\documentstyle[aps,preprint]{revtex}
\include{psfig}
\newcommand{\mywidth}{3.8in}
\begin{document}
\draft
\title{Boundary effects in superfluid films}   
\author{Norbert Schultka$^1$ and Efstratios Manousakis$^2$}
\address{$^1$ Institut f\"ur Theoretische Physik, Technische
Hochschule Aachen, D--52056 Aachen, Germany\\ $^2$Department of Physics, Florida State University, Tallahassee,
Florida 32306, USA}
\date{\today}
\maketitle
\begin{abstract}
We have studied the superfluid density 
and the specific heat of the $x-y$ model on lattices 
$L \times L \times H$ with $L \gg H$ (i.e. on lattices 
representing a film geometry) using the Cluster Monte Carlo 
method. In the $H$-direction we applied staggered
boundary conditions so that the order parameter on the top and 
bottom layers is zero, whereas periodic boundary conditions were 
applied in the $L$-directions. We find that the system exhibits 
a Kosterlitz-Thouless phase transition at the $H$-dependent 
temperature $T_{c}^{2D}$ below the critical temperature 
$T_{\lambda}$ of the bulk system. However, right at the critical 
temperature the ratio of the
areal superfluid density to the critical temperature is $H$-dependent 
in the range of film thicknesses considered here. We do not find
satisfactory finite-size scaling of the superfluid density with 
respect to $H$ for the sizes of $H$ studied.
However, our numerical results can be collapsed onto a single curve by 
introducing an effective thickness $H_{eff} = H + D$ (where $D$ is a
constant) into the corresponding scaling relations.
We argue that the effective thickness depends on the type of 
boundary conditions. Scaling of the specific heat does not require
an effective thickness (within error bars) and we find good agreement 
between the scaling function $f_{1}$ calculated from our Monte Carlo 
results, $f_{1}$ calculated
by renormalization group methods, and 
the experimentally determined function $f_1$.
\end{abstract}
                  
\section{Introduction}
\label{sec1}
The theory of second order phase transitions is based on the assumption that
at temperatures close to the critical temperature $T_{\lambda}$
there is only one dominating length scale associated with the critical 
behavior of the system, the correlation length. Since the correlation
length diverges as the critical temperature is approached the microscopic
details of the system are irrelevant for its critical behavior.
This intuitive picture has its foundation in the renormalization group
treatment of second order phase transitions. Within the renormalization group
treatment it becomes evident that the critical behavior 
can be divided into different universality classes. Each universality class
is characterized by a set of critical exponents which describe the singular
behavior of physical quantities in terms of the reduced temperature
$t=T/T_{\lambda}-1$, e.g. for a three-dimensional bulk system the 
correlation length $\xi(t)$ diverges close to $T_{\lambda}$ as
$\xi(t)=\xi_{0}^{\pm} |t|^{-\nu}$.

If the system is confined in a finite geometry (e.g. a cubic or film geometry)
the singularities in the physical quantities are smoothed out or a crossover
to lower-dimensional critical behavior takes place. 
Finite-size scaling theory \cite{fss} is thought to describe well the 
behavior of the system at temperatures close to $T_{\lambda}$. The intuitive
idea behind the finite-size scaling theory is that finite-size effects
can be observed when the bulk correlation length becomes of the order
of the system size (for a film geometry this is the film thickness $H$).
For a physical quantity $O$ this statement can be expressed as 
follows \cite{brezin}:
\begin{eqnarray}
\frac{O(t,H)}{O(t,H=\infty)}=f\left( \frac{H}{\xi(t,H=\infty)} \right),
 \label{opfss}
\end{eqnarray}
$f$ is a universal function depending only on the geometry and
the boundary conditions applied.

Though earlier experiments on superfluid helium films of
finite thickness \cite{maps} seemed to confirm the validity of the 
approach outlined above, in a recent experiment Rhee, Gasparini, and Bishop 
\cite{rhee,rheephys} showed that their data for the superfluid density 
of thick helium films do not satisfy Eq. (\ref{opfss}) when the expected
value $\nu=0.67$ is used. (For a comprehensive review of experiments on
$^4$He to test the finite-size scaling theory cf. Ref.\cite{FRANCIS}.)
As an attempt to understand these discrepancies 
between theory and experiment 
renormalization group calculations for the standard Landau-Ginzburg 
free energy functional in different geometries with Dirichlet boundary 
conditions (vanishing order parameter at the boundary) 
have been carried out \cite{swdf,dohm,rgdib,HUHN,epsi}. 
New specific heat measurements \cite{lipa} and also a reanalysis
\cite{wado} of the old specific heat  data \cite{earlyc} show good
agreement between the renormalization group  calculations reported in
\cite{swdf,dohm,rgdib,HUHN} and those data. 
These calculations demonstrated the important role played by the boundary
conditions. In particular, periodic boundary conditions were shown to be
inadequate compared to Dirichlet boundary conditions to describe the
experimental specific heat data. 
The renormalization group calculations have determined
the specific heat for that range of
the scaling variable where the surface contribution to the specific
heat is dominant \cite{dohm,rgdib,HUHN,PITZ} (c.f. also \cite{PILA}).
Such field theoretical calculations are not
available for the case of the superfluid density and  the
lack of scaling in the case of the superfluid density of helium films
is not understood. Furthermore, new experiments on
liquid $^{4}He$ under microgravity conditions are planned \cite{lipa2} to 
examine the finite-size scaling properties of the specific heat.
In order to test the renormalization group 
calculations  and because of the reasons above, 
numerical investigations of the 
finite-size scaling properties of the superfluid density \cite{us1} and the
specific heat \cite{us2,us4} of thin helium films have been carried out.
In Refs. \cite{us1,us2}, we used the $x-y$ model with {\it periodic
boundary} conditions in the direction of the film thickness $H$ to 
compute the superfluid density and the specific heat of thin helium films. We 
demonstrated {\it scaling} with respect to the film thickness
using the expected values for the critical exponents 
of the superfluid density and the specific heat, thus confirming the
validity of the finite-size scaling theory. However, the obtained 
universal function for the specific heat does not match the experimentally
determined universal function of Ref. \cite{lipa}, indicating that periodic
boundary conditions are only a poor approximation of the correct
physical boundary conditions as was already demonstrated in Ref.\cite{HUHN}. 
Later we employed staggered-spin boundary 
conditions in the top and 
bottom layers of the film which improves the
agreement between the numerically computed scaling function 
and the experimentally determined scaling function of the
specific heat \cite{us4}.

Another example where the boundary conditions 
play a  role in the scaling behavior comes from Ref. \cite{janna}
where the Villain model, which also belongs to the $x-y$ 
universality class, was studied in a film geometry with open boundary 
conditions in the direction of the film thickness. 
The authors of Ref. \cite{janna} extracted the thickness dependent 
critical temperature from the temperature
dependence of the correlation length in the disordered phase and found
for the critical exponent $\nu$ the value $\nu=0.71(1)$ which is
different from its value of $0.6705$ known from  experiments on liquid 
Helium\cite{golahl}.

In this paper we intend to study the effect of staggered-spin
boundary conditions (Dirichlet--like boundary conditions, i.e. 
vanishing order parameter on the film boundaries)
on the finite-size scaling behavior of 
the superfluid density and the specific heat of $^{4}He$ in a film
geometry in detail. Dirichlet--like boundary conditions
are believed to approximate the physical boundary conditions
more closely \cite{HUHN,Ginzburg}.
Throughout our numerical calculations we are going to describe superfluid
$^{4}He$ near the $\lambda$-critical point by another form of the standard 
Landau-Ginzburg free energy 
functional: the $x-y$ model (cf. e.g. Ref. \cite{kleinert}). In the pseudospin
notation the $x-y$ model takes the following form:
\begin{equation}
{\cal H} = -J \sum_{\langle i,j \rangle} \vec{s}_{i} \cdot 
     \vec{s}_{j}, \label{ham}
\end{equation}
where the summation is over all nearest neighbors, the 
two-component vector $\vec s = (\cos\theta, \sin\theta)$, and 
$J>0$ sets the energy scale. The angle $\theta$ corresponds to the 
phase of the expectation value of the helium atom creation 
operator which is  defined in a volume
whose linear extensions are much larger than the interparticle 
spacing and much smaller than the correlation length. 

In this article we study the superfluid density which corresponds to the
helicity modulus in the pseudospin notation and the
specific heat of the $x-y$ model in a film geometry, 
i.e. on $L^{2} \times H$ lattices with $L \gg H$. The top
and bottom layers are coupled to a static staggered spin 
configuration, playing the role of the ``substrate'' layers, so 
that the magnetization in these layers is exactly zero. 
The crucial difference between these boundary conditions and 
periodic boundary conditions is that the superfluid density 
develops a profile in the $H$-direction, 
whereas it is completely homogeneous for periodic boundary conditions
(cf. also the magnetic profile for the Ising model in a film geometry with
different boundary conditions in Ref. \cite{dohmstau}). We applied periodic
boundary conditions in the $L$-directions because we intend to take the
limit $L \to \infty$.
In the temperature range where the model behaves effectively two-dimensionally
we used the Kosterlitz-Thouless-Nelson renormalization group equations to compute the values for the helicity modulus in the 
$L \rightarrow \infty $ limit. This way we eliminated the 
$L$-dependence of our data for the helicity modulus and were 
able to extract the Kosterlitz-Thouless
transition temperature $T_{c}^{2D}(H)$ for different films.
We investigated the validity of finite-size scaling 
for the superfluid density and the specific heat of such films of
infinite planar dimension and finite $H$. We shall also discuss scaling
of the experimental results for the specific heat and superfluid density
with respect to the film thickness and compare the universal scaling
functions to those obtained from our theoretical investigation.

The rest of this paper is organized as follows. In the next section
we introduce the physical observables defined for the 
$x-y$ model and the numerical method we applied to carry out 
the calculations. In section \ref{sec3}, we discuss the finite-size 
scaling theory and boundary effects.
In section \ref{sec4} we discuss our results 
and the last section briefly summarizes the work described in 
this paper.

\section{Physical observables and Monte Carlo method}
\label{sec2}
For the $x-y$ model on a cubic lattice the helicity modulus is defined as
follows \cite{teiya,litei}:
\begin{eqnarray}
  \frac{\Upsilon_{\mu}}{J} & = & \frac{1}{V} \left\langle 
 \sum_{\langle i,j \rangle} \cos(\theta_{i}-\theta_{j})( \vec{e}_{\mu} \cdot
 \vec{\epsilon}_{ij} )^{2} \right\rangle   \nonumber \\ \protect\nopagebreak
       & & -\frac{\beta}{V} \left\langle \left(
     \sum_{\langle i,j \rangle} \sin(\theta_{i}-\theta_{j}) \vec{e}_{\mu}
   \cdot \vec{\epsilon}_{ij} \right)^{2} \right\rangle ,
 \label{lathel}
\end{eqnarray}
where $V$ is the volume of the lattice, $\beta=J/k_{B}T$, 
$\vec{e}_{\mu}$ is the unit vector in the corresponding bond direction, and 
$\vec{\epsilon}_{ij}$ is the vector connecting the lattice sites $i$ and $j$.
In the following we will omit the vector index since we will always refer to
the $x$-component of the helicity modulus. Note that, because of isotropy,
we have $\Upsilon_{x}=\Upsilon_{y}$. The connection between the
helicity modulus and the superfluid density $\rho_{s}$ is established by  
the relation \cite{fibaja}
\begin{equation}
\rho_s(T) = ( \frac{m}{\hbar} )^{2} \Upsilon(T),
\label{rhos}
\end{equation}
where $m$ denotes the mass of the helium atom.
The specific heat $c$ is obtained from the energy $E$ through
\begin {equation}
 c = \frac{\beta^{2}}{N} 
\left( \langle E^{2} \rangle - \langle E \rangle ^{2} \right),
\end{equation}
where the energy is defined as:
\begin{equation}
 E= \sum_{\langle i,j \rangle} \left( 1- \vec{s}_{i} \cdot 
     \vec{s}_{j} \right)
\end{equation}
and $N$ is the number of spins contributing to the specific heat.

The thermal averages denoted by the angular brackets are computed
according to 
\begin{eqnarray}
\langle O \rangle = Z^{-1} \int \prod_{i} d\theta_{i} \:  O[\theta]
                     \exp( - \beta E ). 
\label{ev}
\end{eqnarray}
$O[\theta]$ denotes the dependence of the physical observable $O$ on the
configuration $\{ \theta_i \}$, the partition function $Z$ is given by
\begin{eqnarray}
  Z=\int \prod_{i} d\theta_{i} \: \exp( -\beta E ), 
 \label{z}
\end{eqnarray}
The multi-dimensional integrals in the  expressions
(\ref{ev}) and (\ref{z}) are computed by means of the Monte Carlo method
using Wolffs 1-cluster algorithm \cite{w1}.

We computed the helicity modulus and the specific heat on 
$L^{2} \times H$ lattices, where $L=20,40,60,80,100$ and $H=4,8,12,16,20,24$.
We applied periodic boundary conditions in the $L$-directions, whereas the
first and the $H$-th layer are coupled to a boundary layer which consists of a
staggered spin configuration, i.e.
\begin{equation}
    \vec{s}(x,y) = (-1)^{x+y}\vec{s}(1,1), \label{stbou}
\end{equation}
where $x,y \in [1,L]$ and label the integer 
coordinates of the lattice 
sites in a plane perpendicular to the $H$-direction. Thus, we have
$V=HL^{2}a^{3}$ with $a$ denoting the lattice spacing and $N=HL^{2}$.
We carried out
of the order of $20,000$ thermalization steps and of the order of
$750,000$ measurements. The calculations were performed on a heterogeneous
environment of computers including Sun, IBM RS/6000 and DEC alpha AXP
workstations and a Cray-YMP.

\section{Finite-size scaling and boundary effects}
\label{sec3}
\subsection{The helicity modulus}
\label{sec3a}
Here we shall discuss the finite-size scaling theory of the superfluid
density. In Ref. \cite{us1}, we studied the helicity modulus 
$\Upsilon$ for the $x-y$ model in a film geometry with {\it periodic} boundary 
conditions in the $H$-direction and we have shown the following 
steps. In a certain temperature range around 
the bulk critical temperature $T_{\lambda}$ where the bulk correlation 
length $\xi(T)$ becomes of the order of the film thickness $H$
the quantity $\Upsilon H/T$
exhibits effectively two-dimensional behavior and 
a Kosterlitz-Thouless phase transition takes place at a temperature 
$T_{c}^{2D}(H) < T_{\lambda}$. 
The critical temperature $T_{c}^{2D}(H)$ approaches
$T_{\lambda}$ in the limit $H \rightarrow \infty$ as 
\begin{equation}
  T_{c}^{2D}(H)=T_{\lambda}\left( 1+\frac{x_{c}}{H^{1/\nu}} \right),
 \label{tch}
\end{equation}
where for periodic boundary conditions we found that 
$x_{c}=-0.9965(9)$ using for the critical exponent $\nu$
the experimental value $\nu=0.6705$ \cite{golahl} and for $T_{\lambda}$
the value $T_{\lambda}=2.2017$ \cite{jantl}.
The quantity $\Upsilon H/T$ is a function of the ratio $H/\xi(T)$, i.e.
\begin{equation}
  \frac{\Upsilon(T,H) H}{T} = \Phi(tH^{1/\nu}).
  \label{tyh}
\end{equation}
The universal function $\Phi(x)$ has the properties \cite{AHNS}
\begin{eqnarray}
  \Phi(x > x_{c}) & = & 0, \nonumber \\
  \Phi(x_{c})     & = & \frac{2}{\pi}.
\end{eqnarray}
In the limit $x \rightarrow x_{c}^-$ the function $\Phi(x)$ can be written as:
\begin{equation}
  \Phi(x)  = \frac{2}{\pi} \left(1 + A \sqrt{x_{c}-x} \right),
 \label{unif} 
\end{equation}
where for periodic boundary condition we found that $A=0.593(5)$. 
This form of the universal function reconciles the
scaling expression (\ref{tyh}) with the two-dimensional behavior
\cite{nelko}, i.e. as $T\to T_{c}^{2D}(H)$
\begin{equation}
 \frac{\Upsilon(T,H) H}{T} =  \frac{2}{\pi}
         \left(1+b(H)\sqrt{1-\frac{T}{T_{c}^{2D}(H)}}\right)
   \label{kinf}
\end{equation}
where \cite{PETSCHEK}
\begin{equation}
  b(H) = A H^{1/2\nu}. \label{bh}
\end{equation}
The results stated above confirm the theoretical expectations about
scaling given
by Ambegaokar et al. in Ref. \cite{AHNS} and agree with the experimental 
findings of Bishop and Reppi \cite{bireppy} and Rudnick \cite{rudnick}. 
In the case of periodic
boundary conditions in the $H$-direction the validity of the 
finite-size scaling form (\ref{tyh}) can already
be observed for films of thicknesses $H=6,8,10$ \cite{us1}.

If nonperiodic boundary conditions are introduced Privman argued that
the general scaling form (\ref{tyh}) has to be altered into \cite{privman}
\begin{equation}
   \frac{\Upsilon(T,H)H}{T} = \bar{\Phi}(tH^{1/\nu}) + \omega \ln H + d,
   \label{yhtdbc}
\end{equation}
where $\omega$ and $d$ are constants depending on the boundary 
conditions. The Monte Carlo data for the helicity modulus
obtained from a Monte Carlo simulation
of the $x-y$ model on cubes $H \times H \times H$ where the 
spins in the boundary layers were all parallel 
(pinned-surface-spin boundary conditions), were found to be
consistent with the presence of the logarithmic term in the 
scaling form (\ref{yhtdbc}) \cite{mon}. 

Let us now investigate the consequences of the logarithmic term 
in (\ref{yhtdbc}).  Again we have to reconcile the two-dimensional
behavior (\ref{kinf}) and the general scaling form (\ref{yhtdbc}) for
temperatures close to the critical temperature $T_{c}^{2D}(H)$ for thick enough
films. Introducing the expression (\ref{tch}) for the $H$-dependent
critical temperature into  Eq. (\ref{kinf})
and keeping only terms up to $H^{1/\nu}$ under the square root leads to
\begin{equation}
  \frac{\Upsilon(T,H)H}{T} = \frac{2}{\pi} + b(H)H^{-1/2\nu}
   \sqrt{x_{c}-tH^{1/\nu}}.
  \label{yhtsca1}
\end{equation}
where we have absorbed the factor of $2/\pi$ in the definition of $b(H)$.
With the assumption (\ref{bh}) we may make the identification
\begin{equation}
  \bar{\Phi}(x)=A \sqrt{x_{c}-x}. \label{phidbc}
\end{equation}
In order to account for the logarithmic term in (\ref{yhtdbc}) we have
to abandon the universal jump at $T_{c}^{2D}(H)$ 
\begin{equation}
  \frac{ \Upsilon(T_{c}^{2D}(H),H) H}{T_{c}^{2D}(H)}=\frac{2}{\pi}, \label{unijum}
\end{equation}
instead we have to assume
\begin{equation}
  \frac{ \Upsilon(T_{c}^{2D}(H),H) H}{T_{c}^{2D}(H)}= d + \omega \ln H,
  \label{hjum}
\end{equation}
i.e. the jump becomes $H$-dependent. (The numerical values for $x_{c}$ and
$A$ will be different from the values given above as they depend on the 
boundary conditions.) 
Eq. (\ref{hjum}) means that for nonperiodic boundary conditions 
in the $H$-direction expression (\ref{kinf}) has to be 
generalized to
\begin{equation}
 \frac{\Upsilon(T,H)H}{T} = g(H) + b(H)\sqrt{1-\frac{T}{T_{c}^{2D}(H)}},
  \label{yht2dg}
\end{equation}
where 
\begin{eqnarray}
  b(H) &=& A H^{1/2\nu}, \label{bhbh} \\
  g(H) &=& d + \omega \ln H. \label{gh}
\end{eqnarray}

\subsection{The specific heat}
\label{sec3b}
For the finite-size scaling of the specific heat one can use similar
scaling expressions to Eq. (\ref{opfss}) which were examined in detail
in Ref.~\cite{us1}. The finite-size scaling expression for the 
specific heat $c$ can also be written in an equivalent way as
\cite{swdf,dohm}:
\begin{equation}
   \left(c(t,H)-c(t_{0},\infty)\right)H^{-\alpha/\nu } = f_{1}(tH^{1/\nu}).
  \label{cimph}
\end{equation}
The function $f_{1}(x)$ is universal and $\nu=0.6705$. At the reduced 
temperature $t_{0}$ the correlation length is equal to the film thickness
$H$, i.e. $t_{0}=(\xi_{0}^{+}/H)^{1/\nu}$ with $\xi_{0}^{+}=0.498$ 
\cite{gottlob}. This scaling form has been used to analyze the
experimental data and, thus, we shall also discuss the scaling of the
specific heat using this form in order to compare to published
experimental results for the universal function $f_1$.
We have
\begin{equation}
c(t_{0},\infty)=c(0,\infty)+\tilde{c}^{+}_{1} \left( \frac{\xi_{0}^{+}}{H}
   \right)^{-\alpha/\nu},
\label{c0l}
\end{equation}
where we have found that $c(0,\infty)=30$, 
$\tilde{c}_{1}^{+}=-30$ \cite{us2} and
$\alpha/\nu=-0.0172$ because of the hyperscaling relation $\alpha=2-3\nu$. 
In Ref. \cite{us2} we demonstrated that our numerical
results for the specific
heat of the $x-y$ model on a film geometry with periodic boundary conditions
follow the finite-size scaling form (\ref{cimph}) for the thicknesses
as small as $H=6,8,10$. 

\section{Results}
\label{sec4}
Here we shall present our results for the superfluid density and the
specific heat and our analysis for the case of 
{\it staggered-spin} boundary conditions as defined in \ref{sec2}.

\subsection{The helicity modulus}

In this section we would like to determine the values of the ratio
$\Upsilon(T,H)H/T$ in the limit $L \rightarrow \infty$ and find estimates
for the critical temperatures $T_{c}^{2D}(H)$ and the parameters $g(H)$ and 
$b(H)$ (cf. Eq. (\ref{yht2dg})). In order to do this we follow closely the
procedure described in Refs. \cite{us1,us3}. 

\begin{figure}[htp] 
 \centerline{\psfig {figure=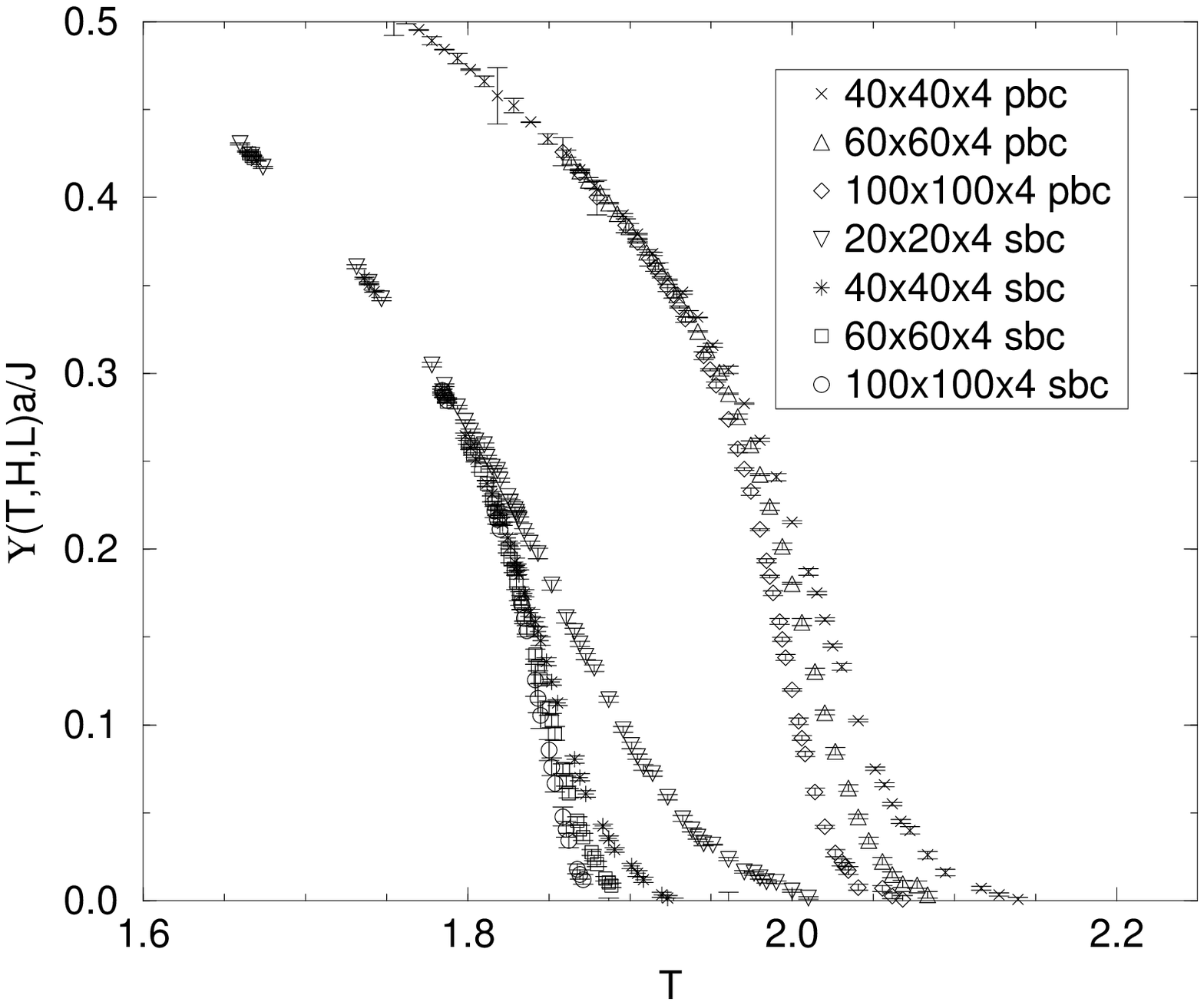,width=\mywidth}}
 \caption{\label{fig0}  The helicity modulus $\Upsilon(T,H,L)$ as
 a function of $T$ for various lattices $L^{2} \times 4$ with staggered
 boundary conditions (sbc) and periodic boundary conditions (pbc) in the
 $H$-direction.}
\end{figure}

Fig.~\ref{fig0} displays the Monte Carlo data for the helicity modulus
in units of the lattice spacing $a$ and the energy scale $J$ for the 
film of fixed thickness $H=4$. This figure demonstrates that
staggered boundary conditions for the top and bottom layers of the film
strongly suppress the values of the helicity modulus with respect to the
case of periodic boundary conditions. As a consequence films with
staggered boundary conditions have a smaller critical temperature
than films with periodic boundary conditions.

\begin{figure}[htp] 
 \centerline{\psfig {figure=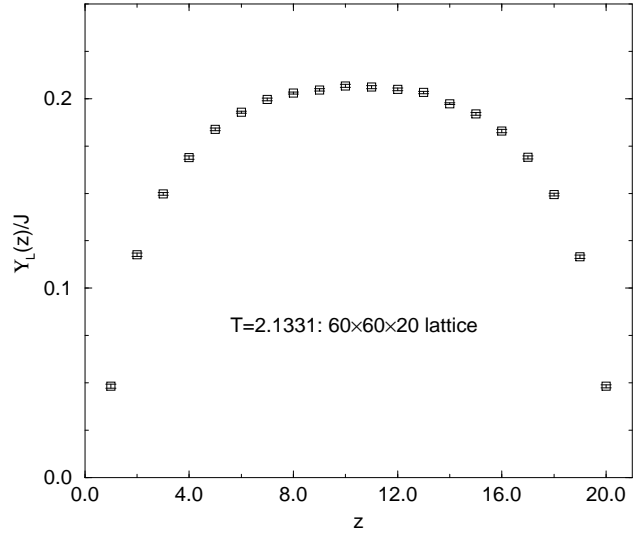,width=\mywidth}}
 \caption{\label{fig0a}  The approximate profile $\Upsilon_{L}(z)$ of the
 helicity modulus computed on a $60 \times 60 \times 20$ lattice at
 $T=2.1331$, i.e. close to the critical temperature, $T_{c}^{2D}(20)=2.1346$.}
\end{figure}

At the temperature $T=2.1331$ we computed the helicity modulus 
on a $60 \times 60 \times 20$ lattice for 
each layer separately and plotted
the result $\Upsilon_{L}(z)/J$ in Fig.~\ref{fig0a}, where $z$ enumerates
the layers. 
The layered helicity  modulus is symmetric with respect to the middle layer 
where it reaches its maximum and decreases when the boundaries are approached. 
Although the helicity modulus $\Upsilon(T,H,L)/J$ is not the average of the
quantity $\Upsilon_{L}(z)/J$ over all layers (this is due to the second
nonlinear term in expression (\ref{lathel})), the curve in Fig.~\ref{fig0a}
is an approximation to
the profile the superfluid density develops in thin films. 

\begin{figure}[htp] 
 \centerline{\psfig {figure=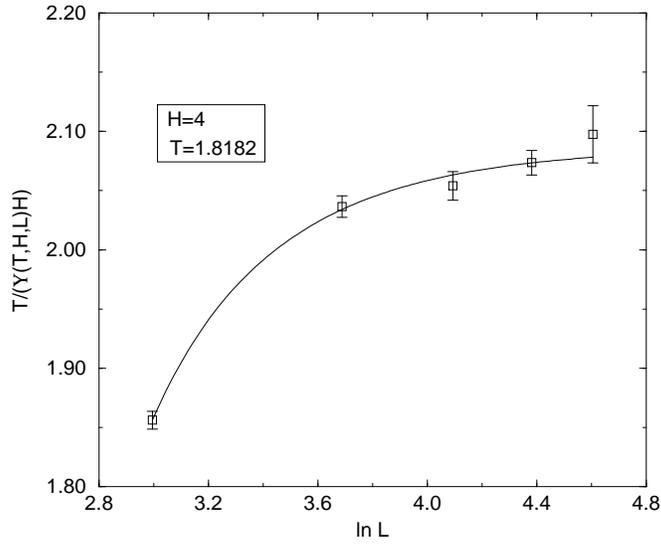,width=\mywidth}}
 \caption{\label{fig1}  $T/(\Upsilon(T,H,L) H)$ as a function
 of $\ln L$ at $T=1.8182$ and $H=4$. The solid curve is the fit to the solution
 to (\protect\ref{dglk}).}
\end{figure}

Let us turn now to the computation of the values for the ratio
$K=T/(\Upsilon H)$ in the $L \rightarrow \infty$ limit.
For a fixed thickness $H$ and 
at temperatures $T$ below but sufficiently close to the critical 
temperature $T_{c}^{2D}(H)$
the system behaves like a two-dimensional system\cite{AHNS,us1}. In this
regime we demonstrated\cite{us1} that
the dimensionless ratio $K$ obeys the 
Kosterlitz-Thouless-Nelson renormalization group equations 
\cite{nelko,jose,THOULESS}:
\begin{eqnarray}
\frac{dK(T,l)}{dl} & = & 4\pi^{3}y^{2}(T,l), \label{rg1} \\
      \frac{dy(T,l)}{dl} & = & (2-\pi K^{-1}(T,l))y(T,l). \label{rg2} 
\end{eqnarray}
$\ln y$ is the chemical potential to create a single vortex, $e^{l}$ denotes
the size of the core radius of a vortex. These equations contain the universal
jump $K(T_{c}^{2D}(H),H)=\pi/2$. In order to adjust the above equations to the
possibility of an $H$-dependent jump of the ratio $K$ at $T_{c}^{2D}(H)$ we 
generalize equations (\ref{rg1}) and (\ref{rg2}) to:
\begin{eqnarray}
\frac{dK(T,l)}{dl} & = & \zeta y^{2}(T,l), \label{rg3} \\
      \frac{dy(T,l)}{dl} & = & 2(1-\epsilon K^{-1}(T,l))y(T,l), \label{rg4} 
\end{eqnarray}
where $\zeta$ and $\epsilon$ are $H$-dependent constants. After eliminating
the variable $y$ from the coupled system of differential equations we
obtain:
\begin{equation}
\frac{dK(T,l)}{dl}=4(K(T,l)-\epsilon \ln K(T,l) - C), \label{dglk}
\end{equation}
where $C$ is a constant which satisfies the condition 
$C \geq \epsilon(1-\ln \epsilon )$. This condition allows for the existence
of roots of the right hand side of Eq. (\ref{dglk}).
If we identify the scale $l$ with $\ln L$ up to a constant we can use
Eq. (\ref{dglk}) to extrapolate the computed values $K(T,H,L)$ 
obtained on lattices of finite planar dimension $L$ to the $L=\infty$
limit. Namely, at $L=\infty$ the left hand side of Eq. (\ref{dglk}) vanishes
\cite{nelko,jose,THOULESS} and $K(T,H,\infty)< \epsilon$ is given by the root 
of the 
right hand side of Eq. (\ref{dglk}). 
The parameters $\epsilon$ and $C$ are found by
fitting the Monte Carlo data for $K(T,H,L)$ at a fixed $H$ to the 
numerical solution to Eq. (\ref{dglk}). 
Table \ref{ta1} contains our fitting results for the
fitting parameters $\epsilon$ and $C$ and the values for $K(T,H,\infty)$
for the thicknesses $H \in [4,20]$ and Fig.~\ref{fig1} shows a typical fit.
We were not able to explore the two-dimensional region for the film with 
$H=24$ because the temperature range where the film behaves two-dimensional
becomes very narrow and our computer resources did not allow an accumulation
of data accurate enough to resolve this region.

\begin{figure}[htp] 
 \centerline{\psfig {figure=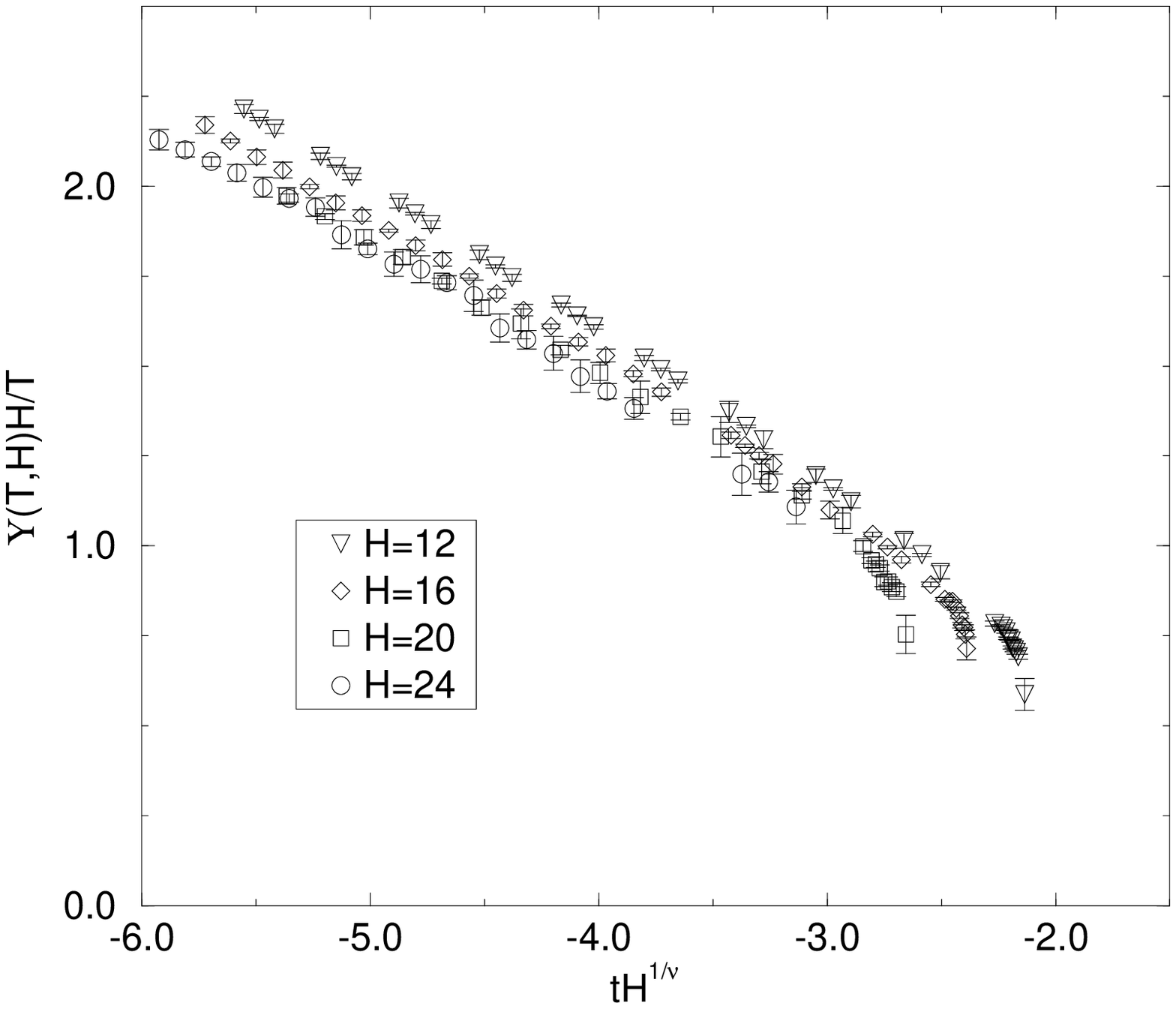,width=\mywidth}}
 \caption{\label{fig1a}  $\Upsilon(T,H) H/T$ as a function of $tH^{1/\nu}$
 for various thicknesses. $\nu=0.6705$.}
\end{figure}

\begin{figure}[htp] 
 \centerline{\psfig {figure=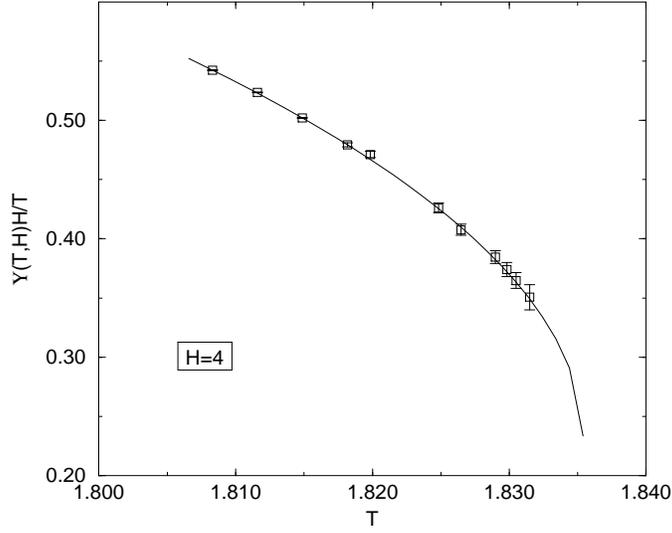,width=\mywidth}}
 \caption{\label{fig2}  $\Upsilon(T,H) H/T$ at $L=\infty$ and $H=4$
 as a function of $T$. The solid curve is the fit to (\protect\ref{yht2dg}).}
\end{figure}

\begin{figure}[htp] 
 \centerline{\psfig {figure=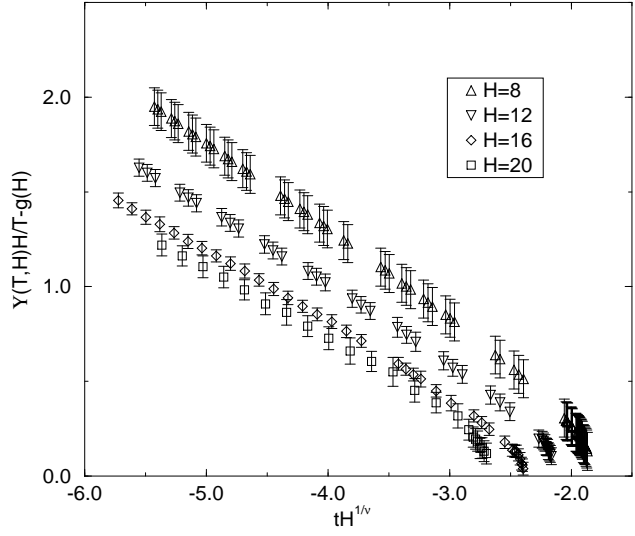,width=\mywidth}}
 \caption{\label{fig3}  $\Upsilon(T,H) H/T-g(H)$ as a function of $tH^{1/\nu}$.
$\nu=0.6705$ and the values for $g(H)$ are taken from Table \protect\ref{ta2}.}
\end{figure}

In Fig.~\ref{fig1a} we plot $\Upsilon(T,H)H/T$ versus $tH^{1/\nu}$ for
the thicknesses $H=12,16,20,24$ to check the validity of 
the scaling form (\ref{tyh}) where $\nu=0.6705$. 
The data for the helicity modulus used in Fig.~\ref{fig1a}
have completely lost their $L$-dependence. We do not obtain a universal 
scaling curve, thus scaling according to the expression (\ref{tyh}) is 
not valid for the films with thicknesses up to $H=24$. Therefore we will try
to employ the scaling form (\ref{yhtdbc}) which requires the knowledge of
$g(H)$. Since $K^{-1}(T,H,\infty)$ satisfies Eq. (\ref{yht2dg}) for a fixed 
$H$ and temperatures close enough to the critical temperature 
$T_{c}^{2D}(H)$ we can fit the
obtained results for $K^{-1}(T,H,\infty)$ to Eq. (\ref{yht2dg}) and find an
estimate for $T_{c}^{2D}(H)$ and the parameters $b(H)$ and $g(H)$. 
In Table \ref{ta2} we present our fitting results and Fig.~\ref{fig2} shows
the fit to the data for $\Upsilon(T,H)H/T$ at $H=4$.
It is interesting to note that the $H$-dependence 
of the parameter $g$ can be described by the formula
\begin{equation}
 g(H)=0.338(19)\ln H -0.238(35) \label{ghnum}
\end{equation}
for $H \in [4,20]$. This is consistent with Privman's prediction 
\cite{privman}.
In Fig.~\ref{fig3} we plot 
$\Upsilon(T,H)H/T-g(H)$ versus $tH^{1/\nu}$ for $H=8,12,16,20$ where $g(H)$ is 
given in Table \ref{ta2}, the bulk critical temperature 
$T_{\lambda}=2.2017$.
Also the scaling form (\ref{yhtdbc}) does not collapse our data points onto 
one universal curve. This situation is the same as the one
Rhee, Gasparini and Bishop encountered when they tried to verify finite-size
scaling for their data of the superfluid density \cite{rhee,rheephys}. 
Their data
of the superfluid density did not fall onto one universal curve when the
scaling form (\ref{tyh}) was employed, neither did the inclusion of a
logarithmic term as in Eq. (\ref{yhtdbc}) help to achieve data collapse
\cite{rheephys}.

Of course, one reason for the failure of scaling of our data of
the helicity modulus according to the expressions (\ref{tyh}) or (\ref{yhtdbc})
could be that our thicknesses are still too small. 
On the other hand Rhee et al.
use films of macroscopic sizes and do not confirm scaling.
Furthermore, for films with periodic boundary conditions scaling of the
helicity modulus occurs already for thicknesses as small as $H=6$ \cite{us1}.

Let us therefore pursue another line of thought \cite{USLOWT} which 
we borrow from the mean field treatment of thin ferromagnetic films 
\cite{KAGANOV}. 
The reduced critical temperature of a ferromagnetic film $t_c(H)$ 
($t_c(H)=1-T_c^{2D}(H)/T_c$ 
where $T_c$ is the 3D bulk critical temperature of the ferromagnet)
can be obtained from the following set of equations \cite{KAGANOV}:
\begin{eqnarray}
u \tan u &=& \frac{H}{2\lambda}, \label{e1} \\
t_c(H) &=& \left(\frac{2a}{H}\right)^2 u^2. \label{e2}
\end{eqnarray}
The lattice spacing is denoted by $a$ and $\lambda$ is the 
extrapolation length $\lambda$ (cf. also Ref. \cite{Binder}). 
Let us compute 
$t_c(H)$ in the limit $H/(2\lambda) \gg 1$. From
Fig.2 of Ref.\cite{KAGANOV} it is clear that $u \rightarrow \pi/2$ in the limit
$H/(2\lambda) \rightarrow \infty$. Thus, writing $u=\pi/2-\epsilon$ with
$ 0< \epsilon \ll 1$ Eq.(\ref{e1}) turns into
\begin{equation}
\frac{2u\lambda}{H}=\tan \epsilon \approx \epsilon =\frac{\pi}{2}-u. \label{e3}
\end{equation}
Solving for $u$ yields
\begin{equation}
u=\frac{\pi}{2\left(1+\frac{2\lambda}{H}\right)}. \label{e4}
\end{equation}
Inserting this into Eq.(\ref{e2}) we obtain finally
\begin{equation}
t_c(H)=\frac{a^2 \pi^2}{(H+2\lambda)^2}. \label{e5}
\end{equation}
Thus, within the mean field treatment the reduced critical temperature
scales with the correct critical exponent $\nu=0.5$ but with an effective
thickness $H+2\lambda$. It is interesting that $2\lambda$ appears in 
Eq.(\ref{e5}). This means that we have to add twice the extrapolation length
(for each side of the film) to the actual thickness.
Furthermore, if the magnetization is suppressed close to the boundaries
$(\lambda > 0)$ the critical temperature $T_c^{2D}(H)$ is smaller than the
3D bulk critical temperature and we have to  add $2\lambda$ to the
actual thickness.

\begin{figure}[htp] 
 \centerline{\psfig {figure=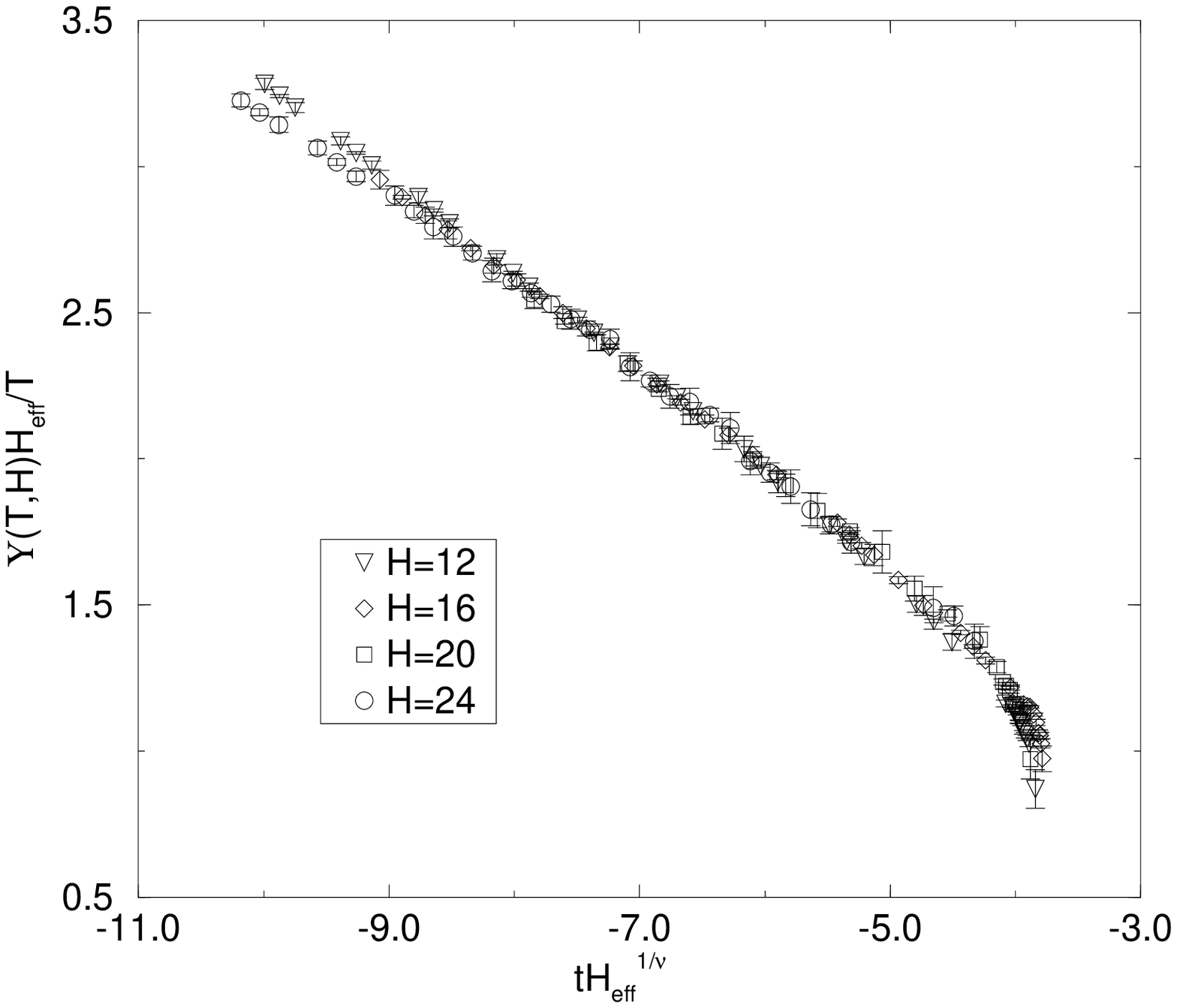,width=\mywidth}}
 \caption{\label{fig4}   $\Upsilon(T,H) H_{eff}/T$ as a function of 
$tH_{eff}^{1/\nu}$ for various thicknesses. $H_{eff}=H+5.79$ and $\nu=0.6705$.}
\end{figure}

The lack of scaling of our data of the helicity modulus 
with the expected critical exponent $\nu=0.6705$ indicates 
that the critical temperatures $T_{c}^{2D}(H)$ do not satisfy Eq. (\ref{tch}).
Instead, due to the profile of the superfluid density  we may expect an
effective film thickness $H_{eff}$ which enters the scaling expressions
(\ref{tch}) and (\ref{tyh}). In close analogy to the mean field treatment of 
ferromagnetic films discussed in the paragraph above, we assume that
$H_{eff}=H+D$ where
$D$ is a constant. Indeed, for the film thicknesses $H=12,16,20$ we obtain
$x_{c}=-3.81(14)$ and $D=5.79(50)$ with $\nu=0.6705$. In Fig.~\ref{fig4}
we plot $\Upsilon(T,H)H_{eff}/T$ as a function of $tH_{eff}^{1/\nu}$ for films
with $H=12,16,20,24$ where $\nu=0.6705$. The data for the helicity 
modulus collapse reasonably well onto a single curve. 
We can understand the
increment $D$ as a scaling correction which renders the scaling 
relations (\ref{tch}) and (\ref{tyh}) valid even for very thin films.
For large thicknesses $H$ the increment $D$ can be neglected and 
we recover the conventional scaling forms. 
Of course, it is possible
to invent scaling forms different from the structure (\ref{e5}) which yield
the conventional scaling expressions in the limit $H \rightarrow \infty$.
For example, we have obtained similarly good fitting results using the
expression $t_c(H)=a_1 H^{-1/\nu}+a_2 H^{-2}$ which is also
motivated by mean field theory (cf. eg. Ref.\cite{RUDNICK}).
However, since we find
Eq.(\ref{e5}) physically appealing we continue to describe the effects
of boundary conditions by an effective thickness as we have done in this
paragraph.

\begin{figure}[htp] 
 \centerline{\psfig {figure=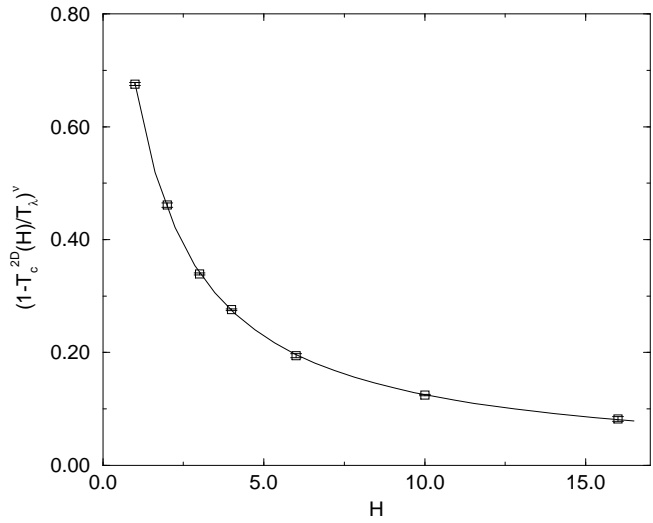,width=\mywidth}}
 \caption{\label{fig5} $(1-T_{c}^{2D}(H)/T_{\lambda})^{\nu}$ for the
Villain model in a film geometry with open boundary conditions as a function 
of $H$. The solid line is the expression (\protect\ref{tchjanke}).}
\end{figure}

In order to test further the assumption that the boundaries 
introduce an effective thickness into the scaling expression 
(\ref{tyh}) we try to describe the thickness dependence
of the Kosterlitz-Thouless transition temperature of the 
Villain model with open boundary conditions 
(interactions of the top and bottom layer only with the 
interior film layers) \cite{janna} by Eq. (\ref{tch}), where $H$
is replaced by the effective thickness $H_{eff}=H+D_{V}$.
Indeed, taking $\nu=0.6705$ we find 
\begin{equation}
  \left(1-\frac{T_{c}^{2D}(H)}{T_{\lambda}}\right)^{\nu}=
  \frac{1.384(9)}{H+1.05(2)}, \label{tchjanke}
\end{equation}
thus $D_{V}=1.05(2)$ and $x_{c}=-1.62(2)$. 
The function (\ref{tchjanke})
is the solid line in Fig.~\ref{fig5}. Again for this case, the
increment $D$ is a correction which makes the scaling relations 
(\ref{tch}) and (\ref{tyh}) valid even for very thin films. 
 The result (\ref{tchjanke}) means that the film thicknesses 
considered in Ref. \cite{janna} were still too small to extract
the expected value of the critical exponent $\nu$ from
the $H$-dependence of the critical temperature (\ref{tch})
without the help of an effective thickness $H+D_{V}$.

\begin{figure}[htp] 
 \centerline{\psfig {figure=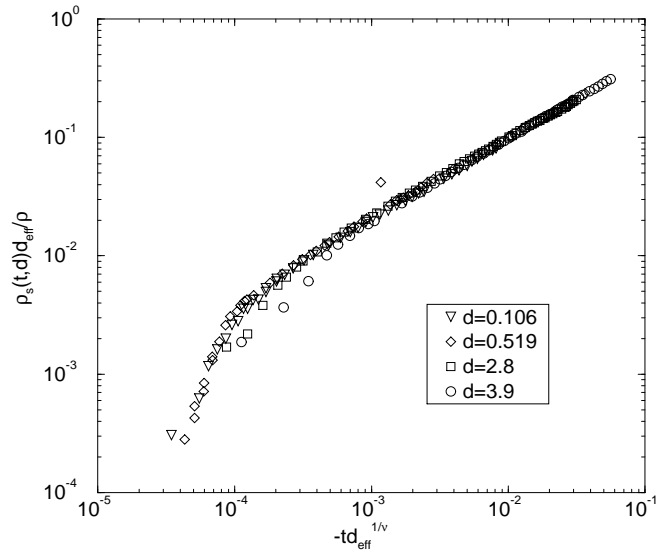,width=\mywidth}}
 \caption{\label{fig6} Scaling of the superfluid density data of Rhee et al.
\protect\cite{rhee} with the effective thickness $d_{eff}=d+0.145$.
$\nu=0.6705$ and all lengths are in $\mu m$.}
\end{figure}

In Fig.~\ref{fig6} we achieve approximate collapse of the experimental data 
of Rhee et al.\cite{rhee,rheephys} for the 
superfluid density $\rho_{s}$ for films of various 
thickness $d$ ($d$ is in $\mu m$) by plotting $\rho_{s}(t,d)d_{eff}/\rho$ 
versus $td_{eff}^{1/\nu}$ with
$\nu=0.6705$ and $d_{eff}=d+0.145$. This value of the effective 
thickness was found by examining the reduced temperatures $t_{fs}(H)$ 
where finite-size effects set in. According to finite-size 
scaling theory $t_{fs}$ has to fulfill the 
relation $t_{fs}\propto d^{-1/\nu}$, thus in our case 
$t_{fs}\propto d_{eff}^{-1/\nu}$. 
The data corresponding to the film with $d=3.9\mu m$
deviate from the universal curve and we attribute this to the 
anomalous behavior of these data. Namely, in general 
$|t_{fs}(d_{1})| > |t_{fs}(d_{2})|$ 
if $d_{1}<d_{2}$, but this is not the case for 
$d_{1}=2.8\mu m$ and $d_{2}=3.9\mu m$ 
(cf. Refs. \cite{rhee,rheephys}). 

Let us compare the increments over the film thickness for the 
three cases of film geometry considered above. 
For the Villain model with open boundary conditions we obtained 
$D_{V}=1.05(2)$ while for the $x-y$ model with staggered boundary 
conditions we obtained  $D=5.8(5)$. All increments are expressed
in lattice spacing units. The difference in these values of the 
increments reflect how severe the effect of the boundary conditions is.
Open boundary conditions are less demanding on the order parameter at the
boundary compared to staggered boundary conditions used in our simulations.
The value of the increment in the case of $^4$He on silicon is
large, $d_{eff}-d=491.5$ in lattice spacing units $a$ 
($a=2.95\mbox{\AA}$ \cite{us2}). 
This might cast some doubts on our proposed scaling form
for the superfluid data of Rhee et al. It is possible, however, to
imagine that on the surface of these films vortices are pinned by
impurities or other forms of disorder, which make the order parameter
vanish at the boundary and which introduce an effective length scale of such a
magnitude. 

So far we have seen that different boundary conditions 
create different effective thicknesses. The influence of the boundaries 
vanishes for thick enough films and only in a certain small range of film
thickness the influence of the boundary conditions has to be taken into
account. According to our findings the scaling form (\ref{unif}) for the
helicity modulus in the limit $T\to T_{c}^{2D}(H)$ takes the 
following form now:
\begin{equation}
   \frac{\Upsilon(T,H) H_{eff}}{T} = \bar{g}
   \left(1 + A \sqrt{x_{c}-tH_{eff}^{1/\nu}} \right),
 \label{unifeff} 
\end{equation}
where $A$, $x_{c}$ and $\bar{g}$ are constants depending on the boundary 
conditions. Especially for $tH_{eff}^{1/\nu}=x_{c}$ we should have 
\begin{equation}
\frac{ \Upsilon(T_{c}^{2D},H) H_{eff} }{ T_{c}^{2D} }=\bar{g}. \label{barg}
\end{equation}
and this value of $\bar g$ is not necessarily equal to $2/\pi$.
In Table \ref{ta3} we give the values for $\bar{g}$ found by our fitting
procedure. We still have a
slight $H$-dependence in $\bar{g}$ but for the thicknesses $H \geq 16$ 
the value for $\bar{g}$ seems to saturate at $\bar{g}(H\to \infty)
 \approx 0.97$.
Since we can neglect the effective thickness for very large film thicknesses,
this means that films with staggered boundary conditions in the 
$H$-direction of the film exhibit a jump in the quantity 
$\Upsilon(T_{c}^{2D},H)H/T_{c}^{2D}$ that is different from $2/\pi$ which
was found for films with periodic boundary conditions \cite{nelko,jose,us1}.
Therefore, assuming that our extrapolation to large film
thicknesses from small size films using the idea of the effective
thickness is valid, we have to conclude that the jump 
$\Upsilon(T_{c}^{2D},H)H/T_{c}^{2D}$ depends on the boundary conditions.
In principle there is nothing wrong with this conclusion because the
universal functions (and the jump is a particular feature of a 
particular universal function) depend very importantly (especially near
the critical temperature) on the boundary conditions. The scaling
function should not be confused with the critical exponents which are
independent of the geometry and boundary conditions. The scaling
functions for given universality, given geometry and boundary conditions
are universal. This leads us to the conclusion that this jump in the
experimental findings should depend on the substrates 
(cf. also Ref.\cite{USLT21}). 
This influence of the substrate must be mediated by the vortices
whose generation is enhanced close to the boundaries due to the effect of the
boundary (for a more detailed discussion cf. section \ref{disc}). 
In the experiments the value of $2/\pi$ was found 
\cite{bireppy,rudnick}, however, the thicknesses of these films are much
smaller than the above length scale $D$ found to fit the data of Rhee et
al. We believe that 
the vortex density was almost homogeneous throughout
the films used for these 
measurements. This situation corresponds to the $x-y$ model in a film geometry
with periodic boundary conditions along the film direction where the vortex
density is the same everywhere.

\subsection{The specific heat}

\begin{figure}[htp] 
 \centerline{\psfig {figure=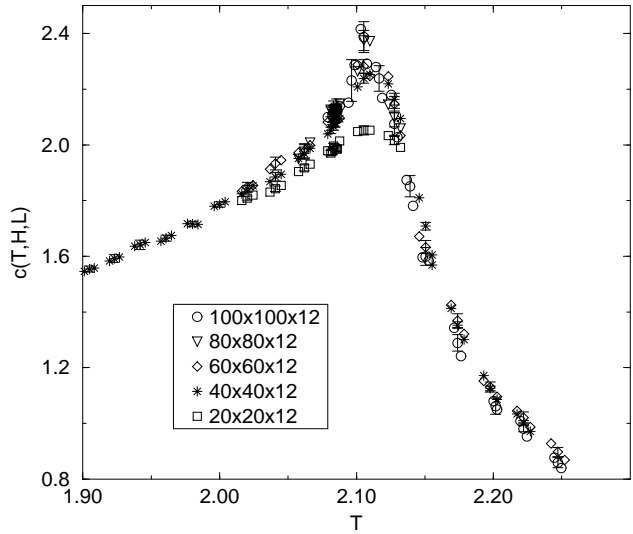,width=\mywidth}}
 \caption{\label{fig8} The specific heat $c(T,H,L)$ as a function of $T$
  for $L^{2}\times 12$ lattices. $T_{c}^{2D}(12)=2.086$, $T_{\lambda}=2.2017$.}
\end{figure}

\begin{figure}[htp] 
 \centerline{\psfig {figure=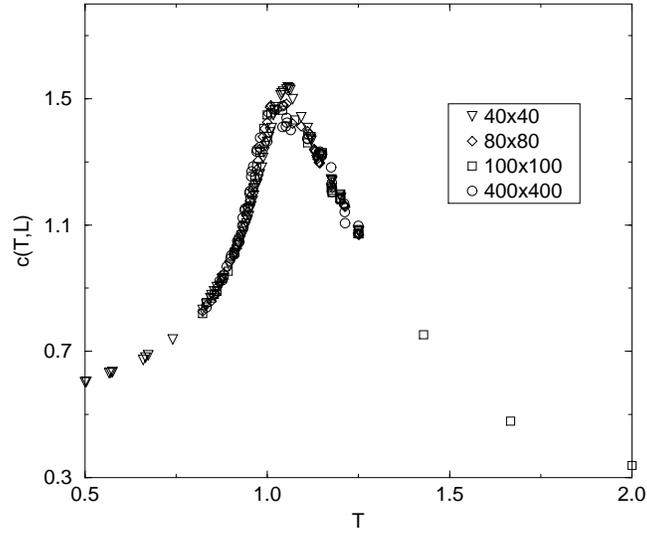,width=\mywidth}}
 \caption{\label{fig11} The specific heat $c(T,L)$ for pure two-dimensional 
 lattices $L \times L$ with periodic boundary conditions.}
\end{figure}

In this section we would like to investigate the finite-size behavior
of the specific heat $c(T,H)$. 
Since we do not possess an easily handable procedure to take the 
$L \rightarrow \infty$ limit for the values of the specific heat $c(T,H,L)$
computed on finite lattices $L\times L \times H$ we approximate films
with infinite planar dimension by $100 \times 100 \times H$ lattices. 
This seems justified because the specific heat appears independent of $L$
for $L \geq 60$ (cf. Fig.~\ref{fig8}).
Furthermore, we do not expect the maximum of the specific heat to grow 
dramatically with increasing values of $L$ because for temperatures in the 
range $T_{c}^{2D}(H)\leq T \leq T_{\lambda}$ the behavior of the specific heat
can be described by the Kosterlitz-Thouless theory which leads to a finite
value of this maximum. In order to illustrate this argument we show in 
Fig.~\ref{fig11} the size dependence of the specific heat $c(T,L)$ computed on
pure two-dimensional lattices $L \times L$ with periodic boundary conditions.
The $L$-dependence of the specific heat can be neglected for 
values of $L>80$.

\begin{figure}[htp] 
 \centerline{\psfig {figure=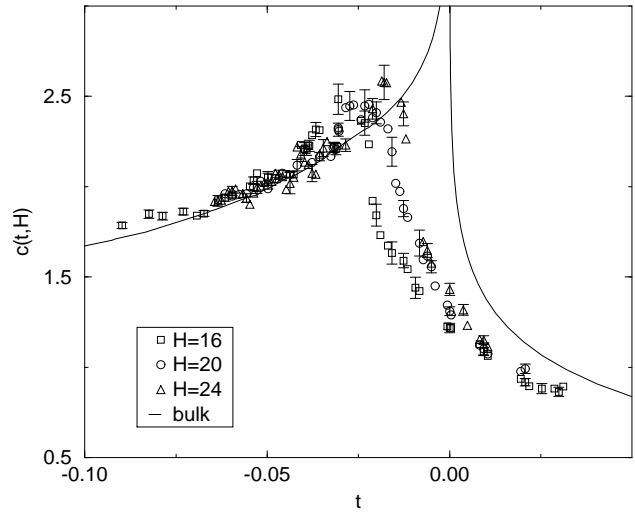,width=\mywidth}}
 \caption{\label{fig13a} The specific heat $c(t,H)$ for various films of 
 finite thickness $H$ compared to the bulk specific heat $c(t,\infty)$. }
\end{figure}

\begin{figure}[htp] 
 \centerline{\psfig {figure=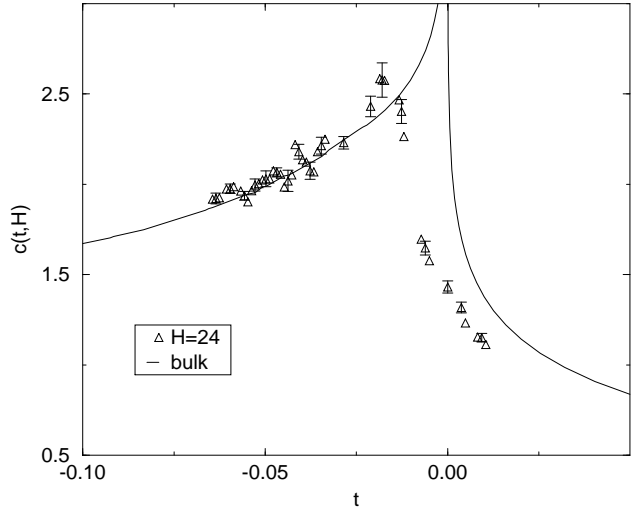,width=\mywidth}}
 \caption{\label{fig13b} The specific heat $c(t,H)$ for the film of 
  thickness $H=24$ compared to the bulk specific heat $c(t,\infty)$.}
\end{figure}

In Fig.~\ref{fig13a} we compare the specific heat of films with $H=16,20,24$
to the bulk specific heat $c(t,H=\infty)$ taken from Ref. \cite{us2}.
The specific heat values for films of thickness $H=12,16,20$ lie above the
bulk curve for $t<0$.
Such a behavior is also found in experiments on
helium films about 30\AA{} thick \cite{FRANCIS}. This crossing effect is due
to the large shift of the temperature $T_m$ where the specific 
heat for a certain film thickness takes its maximum
down to temperatures below the bulk critical
temperature $T_{\lambda}$. For thicker and thicker films
the maximum temperature $T_m$ approaches $T_{\lambda}$ , thus the confined
specific heat data will fall below the bulk curve which is expected from the
field theoretical calculations\cite{dohm,rgdib,HUHN,DOHMEXP}. 
The film thicknesses used in our Monte Carlo calculation range from 
$H=35\mbox{\AA}$ to $H=70.8\mbox{\AA}$ 
(the lattice spacing $a=2.95\mbox{\AA}$ \cite{us2})
which is comparable to the experimental film thicknesses (30\AA) where this
crossing effect can be observed.
Fig.~\ref{fig13b} shows the specific heat of the film with $H=24$
alone, indicating that the effect of crossing the bulk curve 
indeed vanishes for thicker and thicker films.
Unfortunately it is beyond our means to carry out the necessary
analysis for thicker films than were treated here.
It is interesting to note that we find the same qualitative 
behavior of the specific heat in the case of the $x-y$ model in a cylindrical 
geometry \cite{uspores}.

\begin{figure}[htp] 
 \centerline{\psfig {figure=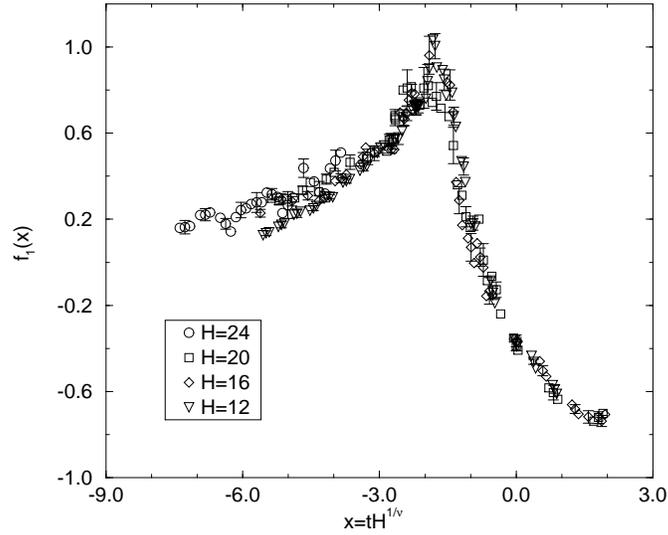,width=\mywidth}}
 \caption{\label{fig8b} Scaling function $f_{1}(x)$ for films with 
 staggered boundary conditions where $x=tH^{1/\nu}$.}
\end{figure}

In Fig.~\ref{fig8b} we plot the scaling function $f_{1}(x)$ given by expression
(\ref{cimph}). According to the previous discussion this scaling
function can only be 
an approximation to the correct one which one would need to compute from
films with $H>24$ and $L >> H$. 
We find that our data for the specific heat for films of various thicknesses
collapse approximately onto a single curve. 
It seems that the specific heat is rather insensitive to the boundary effect 
of introducing an effective thickness and a very high accuracy in the 
computation of the specific heat is needed to detect the presence of the
effective thickness in the scaling function $f_{1}(x)$. 
For example, one could determine the temperatures $T_{m}(H)$ where the 
specific heat reaches its maximum and examine the validity of Eq. (\ref{tch})
for $T_{m}(H)$, because the $H$-dependence of $T_{m}(H)$ is also given by 
expression (\ref{tch}) 
(with a different value for $x_{c}$ than for the critical temperatures 
$T_{c}^{2D}(H)$). This can be done more easily in experiments because 
in Monte Carlo calculations an extrapolation procedure for the values of
the specific heat at finite planar dimensions to the values at infinite
planar dimensions is needed (and which is not available at present) whereas
the films used in experiments represent films with infinite planar dimensions.

\begin{figure}[htp] 
 \centerline{\psfig {figure=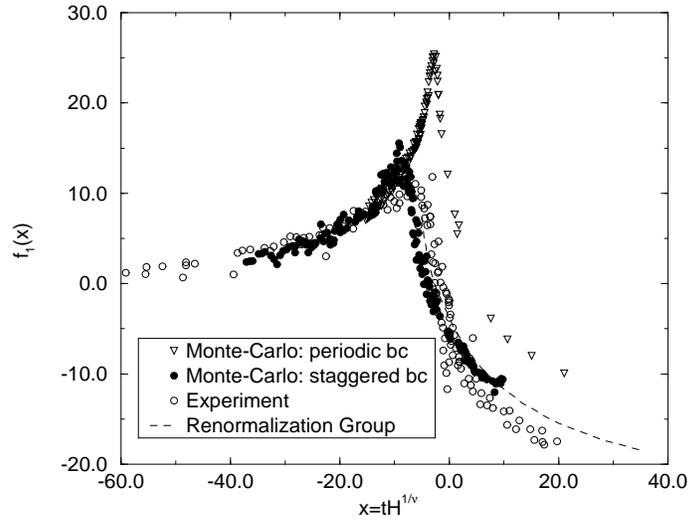,width=\mywidth}}
 \caption{\label{fig12} The experimentally determined function
 $f_{1}(x)$ (open circles) of Refs. \protect\cite{lipa}, $f_{1}(x)$ 
 for films with periodic boundary conditions (triangles) of 
 Ref. \protect\cite{us2} and staggered boundary conditions (filled circles), 
 and  $f_{1}(x)$ for films with Dirichlet boundary conditions of Ref.
 \protect\cite{dohm} (dashed line).}
\end{figure}

We can directly compare our function $f_{1}(tH^{1/\nu})$ to the 
experimentally determined scaling function $f_{1}(x)$ given in Refs. 
\cite{lipa} by expressing all lattice units in physical
units using the conversion formula:
\begin{equation}
\left.f_{1}(x) \right|_{phys}=\frac{V_{m}k_{B}}{a^{3}} \left(
      \frac{a}{\mbox{\AA}}\right)^{-\alpha/\nu}
      \left.f_{1}(x)\right|_{lattice}=15.02 \frac{\mbox{Joule}}{
^{\circ}\mbox{K mole}} \left.f_{1}(x)\right|_{lattice}, \label{convert}
\end{equation}
where $V_{m}$ is the molar volume of $^{4}He$ at saturated vapor pressure
at $T_{\lambda}$, the lattice spacing $a=2.95\mbox{\AA}$ \cite{us2} and
the film thickness is measured in \AA. 
In Fig.~\ref{fig12} we compare the functions $f_{1}(x)$ obtained from 
Monte Carlo calculations of the specific heat of films with periodic
boundary conditions and staggered boundary conditions in the direction
of the film thickness to the experimentally determined function $f_{1}(x)$
and to the function $f_{1}(x)$ obtained from field theoretical calculations
\cite{dohm,rgdib}. 
This figure clearly shows the influence of the boundary conditions
on the shape of the universal function as was already demonstrated by the 
field theoretical calculations reported in Ref. \cite{HUHN}. 
In Fig.~\ref{fig12} we see that 
the scaling function $f_{1}(x)$ for films with staggered boundary conditions
crosses  the scaling function $f_{1}(x)$ for films with periodic boundary
conditions (cf. also \cite{us4}) in the range
$-14 < tH^{1/\nu} < -8$ (cf. Fig.~\ref{fig12}) with $H$ measured in
\AA. This crossing is due to the relative smallness of our film thicknesses
(see the discussion above) and does not occur if we had used much thicker
films in our Monte Carlo calculations to deduce the scaling function $f_1(x)$
\cite{DOHMEXP}.
We expect our function $f_1(x)$ to be slightly modified in the range
$-14 < tH^{1/\nu} < -8$ (cf. Fig.\ref{fig12}) when it is computed from much 
thicker films. We believe that the wings of our curve, however, 
will remain unchanged.
Unfortunately, it is beyond our computational means to repeat the calculations
for thicknesses larger than 24.

\section{Discussion of the results}
\label{disc}
Our findings suggest that it is possible to introduce an effective
thickness $H_{eff}=H+D$ into the scaling expressions for the 
superfluid density (\ref{tyh})  
and achieve scaling (i.e. data collapse) even for rather
thin films. The increment $D$ over the film thickness $H$ can be understood
as an effective correction to scaling. The appearance of the effective 
thickness in our scaling-function can be understood as follows.
The superfluid density has a finite value in the first layers next to
the boundary layers of the film, and its
rise from this finite value to its bulk value $\rho_s^m(T,H)$ 
(for $T < T_C^{2D}(H)$)  inside the film can be divided
into two regions. Let us assume that $T$ is very close to $T_c^{2D}(H)$ 
where the correlation length is very large.
There is a rather narrow region of thickness $D_1$ of the film  which is in
contact with the boundary wall where the superfluid density rises very
fast to attain some value $\rho_s^{D_1}<\rho_s^m$. Then it
rises with a much slower rate over a length scale of the order of the
correlation length to reach its value of $\rho_s^m$. 
The reason for the initial fast rise are the correlations over length
scales much smaller than the correlation length.
One might think that one then has to exclude the region
of the film where the superfluid density rises very sharply
and this leads to a negative value for $D$. However, 
this initial rise of $\rho_s$ occurs very fast over a length scale
$D_1$ which is smaller than a length $D_2$ which would have been
required in order for the superfluid density to reach the same value 
if this rise would have occurred over larger distances over which
the spin-spin correlations are governed by the correlation
length which controls the long distance behavior of the 
correlation function. This implies that the required increment to the
thickness is $D = D_2 -D_1$, which is positive. 

Only for films with thicknesses $H$ which fulfill $H \gg D$ 
scaling with $H$ can be observed.
For periodic boundary conditions we have $D=0$ \cite{us1}, open boundary
conditions seem to yield $D=1.05$ and for staggered boundary conditions we
obtain $D=5.79$. Due to their structure staggered boundary conditions support 
vortex formation close to the boundaries, thus the superfluid density 
decreases from its maximum in the middle of the film towards the boundaries. 
This effect is less pronounced for open and absent for periodic boundary 
conditions. Our results imply that the more the superfluid 
density is suppressed
near the boundaries the larger is the value of $D$. The data for the 
superfluid density of Rhee et al. \cite{rhee} which correspond to $^{4}He$
on Si require  a large value of $D=0.145 \mu m$. Thus, 
Si should suppress the superfluid density dramatically close to the boundary. 
Since $D$ is so large, only films with $H > 3\mu m$ should allow scaling
with $H$. It would be interesting to investigate the scaling behavior
of the superfluid density of $^{4}He$ on different substrates (which represent
different types of boundary conditions) in a wide range 
of film thicknesses to check our hypothesis. 

A consequence of scaling our data of the helicity modulus (or superfluid
density) using an effective thickness is that the jump in the 
quantity $\Upsilon(T_{c}^{2D},H)H/T_{c}^{2D}$ depends on the boundary
conditions in the top and bottom layers of the film, i.e. on the substrates
in real helium experiments. The value of the jump
is only $2/\pi$ (in lattice units) for film thicknesses 
small compared to $D$ as is the case in the experiments reported 
in Refs. \cite{bireppy,rudnick}. The same value of the jump was found is the 
case for $x-y$ films with periodic boundary
conditions \cite{us1}. Thus, experiments could be also used to determine
the jump in the areal superfluid density at the 
Kosterlitz-Thouless temperature and determine its substrate dependence.

In this work we represented the substrate by a staggered spin configuration 
coupled to the top and bottom layers of the film to simulate 
Dirichlet-like boundary conditions 
(zero order parameter in the boundary) in the substrate.
For the staggered spin configuration the local magnetization is exactly zero 
on a plaquette, i.e. on a domain of the size $a \times a$ where $a$ denotes
the lattice spacing. Dirichlet-like boundary conditions in the substrate are,
however, also realized by the following spin configuration:
The spins are parallel in a square domain of linear dimension $R$, but any
two spins representing two adjacent domains are antiparallel to each other.
Thus, an additional length scale associated with a finite
value of the local magnetization over the length scale $R$ is introduced and 
influences the scaling behavior of the helicity modulus and the specific heat.
Since disorder in the 
boundaries supports vortex formation close to the boundaries, vortices
should play an active role in creating the boundary effect described above.

\section{Summary}
\label{sec5}
We have investigated the finite-size scaling properties of the specific heat
$c$ and the helicity modulus $\Upsilon$ of the $x-y$ model in a film geometry,
i.e. on $L \times L \times H$ lattices with $L \gg H$ where
staggered and periodic boundary conditions where applied in the $H$-direction
and the $L$-directions of the film, respectively. We found that a 
Kosterlitz-Thouless phase transition takes place at the $H$-dependent 
critical temperatures $T_{c}^{2D}(H)$, however, for the films used in our
calculations the jump $\Upsilon(T_{c}^{2D}(H),H)H/T_{c}^{2D}(H)$ appears to
be $H$-dependent. Furthermore, scaling of the helicity modulus according
to Eq. (\ref{tyh}) is not valid for our film thicknesses, 
neither is scaling according to Eq. 
(\ref{yhtdbc}) which was derived following a suggestion of 
Privman \cite{privman}. Introducing an effective thickness
$H_{eff}>H$ into the scaling expression (\ref{tyh}) we are able to collapse 
our data as well as the data of Rhee et al. \cite{rhee} reasonably well
onto one universal curve. Our results suggest that the boundary effect of
creating an effective thickness $H_{eff}$
depends on the boundary conditions which can be realized in experiments by
different substrates and is negligible for thicknesses $H$ which fulfill
$H_{eff}/H-1\ll 1$. We argue that the jump in the quantity
$\Upsilon(T_{c}^{2D},H)H/T_{c}^{2D}$ depends on the boundary conditions
and is $2/\pi$ only for certain ideal boundary conditions such as the
periodic boundary conditions.
Within error bars scaling of the specific heat does not require an effective 
thickness and the scaling function $f_{1}(x)$ for the specific heat 
agrees rather well with the experimentally determined 
scaling function $f_{1}(x)$ and with the result of the renormalization
group calculations reported in \cite{dohm,rgdib}. However, we found that 
Monte Carlo simulations of much thicker films than we have used have to be 
performed to determine the position of the maximum of the scaling function 
accurately. At present this is unfortunately beyond our computer resources.

\section{Acknowledgements}
We would like to thank Prof. Dohm for interesting and useful discussions.
N. S. would like to thank the H\"ochsleistungsrechenzentrum J\"ulich for 
the opportunity of using their computing facilities.
This work was supported by the National Aeronautics and Space
Administration under grant no. NAG3-1841 and by Sonderforschungsbereich
341 der Deutschen Forschungsgemeinschaft.

\begin{table}[htp] \centering
 \begin{tabular}{|l|l|l|l|l|l|l|} 
 \hline
 \multicolumn{1}{|c|}{$H$}  & \multicolumn{1}{c|}{$T$} &
 \multicolumn{1}{c|}{$\epsilon$} & \multicolumn{1}{c|}{$C$} &
 \multicolumn{1}{c|}{$K(T,H,\infty)$} &\multicolumn{1}{c|}{$\chi^{2}$} &
 \multicolumn{1}{c|}{$Q$} \\ 
\hline\hline
4 & 1.8315 & 3.2080(6) & -0.51(1)  & 2.852(87) & 0.74 & 0.39 \\
  & 1.8305 & 3.2251(4) & -0.511(9) & 2.742(50) & 0.16 & 0.69 \\
  & 1.8298 & 3.2222(5) & -0.495(9) & 2.674(43) & 0.17 & 0.68 \\
  & 1.8290 & 3.2125(5) & -0.470(9) & 2.601(37) & 0.15 & 0.70 \\
  & 1.8265 & 3.0255(7) & -0.262(7) & 2.453(29) & 0.58 & 0.45 \\
  & 1.8248 & 3.0353(8) & -0.243(7) & 2.347(23) & 0.45 & 0.50 \\
  & 1.8198 & 3.1442(6) & -0.244(6) & 2.122(12) & 0.02 & 0.89 \\
  & 1.8182 & 3.0890(7) & -0.186(3) & 2.0866(72)& 0.66 & 0.52 \\
  & 1.8149 & 3.5845(1) & -0.478(8) & 1.9916(9) & 0.03 & 0.86 \\
  & 1.8116 & 5.21092(2)& -1.46(1)  & 1.9094(7) & 0.009& 0.92 \\
  & 1.8083 & 10.53880(4)& -4.60(4) & 1.8434(8) & 0.06 & 0.81 \\ \hline
8 & 2.0167 & 2.0420(2) & 0.624(3)  & 1.666(14) & 1.76 & 0.18 \\
  & 2.0161 & 2.0787(2) & 0.614(3)  & 1.633(12) & 1.63 & 0.20 \\
  & 2.0155 & 2.1046(2) & 0.608(3)  & 1.608(11) & 1.58 & 0.21 \\
  & 2.0147 & 2.0465(1) & 0.641(2)  & 1.5890(84)& 0.16 & 0.69 \\
  & 2.0141 & 2.0371(1) & 0.650(2)  & 1.5750(82)& 0.12 & 0.73 \\
  & 2.0135 & 2.0262(1) & 0.659(2)  & 1.5610(83)& 0.07 & 0.79 \\
  & 2.0127 & 2.41711(9)& 0.509(3)  & 1.5190(57)& 3.12 & 0.08 \\
  & 2.0121 & 2.50265(9)& 0.483(4)  & 1.5032(53)& 3.20 & 0.07 \\
  & 2.0115 & 2.53788(9)& 0.477(4)  & 1.4908(57)& 3.08 & 0.08 \\
  & 2.0107 & 4.78735(2)& -0.383(9) & 1.4739(42)& 1.56 & 0.21 \\
  & 2.0101 & 5.54710(2)& -0.65(1)  & 1.4629(42)& 1.48 & 0.22 \\ 
  & 2.0094 & 3.39908(6)& 0.183(6)  & 1.4528(45)& 1.44 & 0.23 \\ \hline
12& 2.0846 & 1.7021(3) & 0.818(3)  & 1.446(14) & 1.17 & 0.28 \\
  & 2.0842 & 1.7190(2) & 0.817(3)  & 1.420(12) & 0.77 & 0.38 \\
  & 2.0838 & 1.7148(3) & 0.822(3)  & 1.406(12) & 0.52 & 0.47 \\
  & 2.0833 & 1.7221(1) & 0.822(2)  & 1.3949(86)& 0.25 & 0.78 \\ 
  & 2.0829 & 1.7664(2) & 0.817(2)  & 1.3582(64)& 1.34 & 0.26 \\   
  & 2.0825 & 1.7455(2) & 0.826(2)  & 1.3500(66)& 1.32 & 0.27 \\
  & 2.0822 & 1.7343(2) & 0.831(2)  & 1.3466(68)& 1.29 & 0.28 \\
  & 2.0818 & 2.1259(1) & 0.733(3)  & 1.3140(48)& 0.004& 0.95 \\ 
  & 2.0812 & 2.1171(1) & 0.745(3)  & 1.2998(47)& 0.007& 0.93 \\
  & 2.0805 & 2.1161(1) & 0.754(3)  & 1.2859(49)& 0.06 & 0.81 \\ \hline
16& 2.1173 & 1.4971(3) & 0.903(3)  & 1.327(21) & 1.36 & 0.24 \\
  & 2.1171 & 1.5517(3) & 0.894(3)  & 1.294(14) & 1.56 & 0.21 \\
  & 2.1169 & 1.5591(3) & 0.895(3)  & 1.282(13) & 1.53 & 0.22 \\
  & 2.1164 & 1.5589(3) & 0.904(3)  & 1.241(12) & 0.14 & 0.87 \\ 
  & 2.1160 & 1.7307(2) & 0.880(3)  & 1.2104(61)& 1.64 & 0.19 \\
  & 2.1153 & 2.0342(4) & 0.842(3)  & 1.1825(48)& 1.05 & 0.35 \\
  & 2.1148 & 1.7092(3) & 0.896(2)  & 1.1816(56)& 2.20 & 0.11 \\
  & 2.1142 & 1.5761(3) & 0.921(2)  & 1.1748(68)& 0.40 & 0.67 \\
  & 2.1119 & 1.6254(5) & 0.936(4)  & 1.1197(78)& 0.42 & 0.66 \\ \hline
20& 2.1336 & 1.280(1)  & 0.972(2)  & 1.146(18) & 0.32 & 0.73 \\
  & 2.1331 & 1.273(2)  & 0.974(2)  & 1.130(16) & 0.45 & 0.64 \\
  & 2.1327 & 1.273(3)  & 0.977(2)  & 1.109(14) & 0.55 & 0.58 \\
  & 2.1322 & 1.238(2)  & 0.981(2)  & 1.111(17) & 0.13 & 0.88 \\
  & 2.1317 & 1.310(3)  & 0.982(2)  & 1.066(10) & 0.05 & 0.95 \\
  & 2.1313 & 1.303(2)  & 0.986(2)  & 1.0535(94)& 0.04 & 0.96 \\
  & 2.1308 & 1.293(1)  & 0.989(2)  & 1.0428(91)& 0.06 & 0.94 \\
  & 2.1299 & 1.692(2)  & 1.00(1)   & 1.001(14) & 0.03 & 0.86 \\ 
 \hline \hline
 \end{tabular}
\caption{\label{ta1} Fitted values of the parameters $\epsilon$ and $C$ of
the expression (\protect\ref{dglk}) and the extrapolated values 
$K(T,H,\infty)$ at various temperatures $T$ and different thicknesses $H$.
 $\chi^{2}$ and the goodness of the fit $Q$ are also given.}
\end{table}

\begin{table}[htp] \centering
 \begin{tabular}{|l|l|l|l|l|l|} 
 \hline
 \multicolumn{1}{|c|}{$H$}  & \multicolumn{1}{c|}{$g(H)=\Upsilon(T_{c}^{2D}(H),H)H/T_{c}^{2D}(H)$} & 
 \multicolumn{1}{c|}{$b(H)$} &  \multicolumn{1}{c|}{$T_{c}^{2D}(H)$} & 
 \multicolumn{1}{c|}{$\chi^{2}$} &  \multicolumn{1}{c|}{$Q$} \\ 
\hline\hline
4  & 0.231(13) & 2.564(63) & 1.8354(9) & 1.15 & 0.33 \\
8  & 0.47(10)  & 2.94(83)  & 2.0207(41)& 0.88 & 0.54 \\
12 & 0.587(44) & 3.68(57)  & 2.0862(10)& 1.32 & 0.24 \\
16 & 0.715(32) & 4.07(78)  & 2.1175(3) & 0.23 & 0.87 \\
20 & 0.754(53) & 4.91(91)  & 2.1346(7) & 0.73 & 0.60 \\
 \hline
 \end{tabular}
\caption{\label{ta2} Fitted values of the parameters $g(H)$, $b(H)$ and 
$T_{c}^{2D}(H)$ of the expression (\protect\ref{yht2dg}) for different 
thicknesses $H$.
$\chi^{2}$ and the goodness of the fit $Q$ are also given.}
\end{table}

\begin{table}[htp] \centering
 \begin{tabular}{|l|l|} 
 \hline
 \multicolumn{1}{|c|}{$H$}  & \multicolumn{1}{c|}{$\bar{g}=\Upsilon(T_{c}^{2D}(H),H)H_{eff}/T_{c}^{2D}(H)$} \\
\hline\hline
4  & 0.565(32) \\
8  & 0.81(17)  \\
12 & 0.870(65) \\
16 & 0.974(44) \\
20 & 0.972(68) \\
 \hline
 \end{tabular}
\caption{\label{ta3} The jump
$\bar{g}=\Upsilon(T_{c}^{2D}(H),H)H_{eff}/T_{c}^{2D}(H)$ for different 
thicknesses $H$.}
\end{table}

\end{document}